\documentclass[11pt,twoside]{article}
\usepackage{asp2010}
\usepackage{graphicx}

\resetcounters

\begin{document}

\title{Simulations of binary galaxy mergers and the link with Fast Rotators, Slow Rotators, and Kinematically Distinct Cores}
\author{Maxime Bois$^{1}$, Eric Emsellem$^{2}$, Fr\'ed\'eric Bournaud$^{3}$, Katherine Alatalo$^4$, Leo Blitz$^4$, Martin Bureau$^5$, Michele Cappellari$^5$, Roger L. Davies$^5$, Timothy A. Davis$^5$, P. T. de Zeeuw$^{2,6}$, Pierre-Alain Duc$^{3}$, Sadegh Khochfar$^7$, Davor Krajnovi\'c$^2$, Harald Kuntschner$^{8}$, Pierre-Yves Lablanche$^{2}$, Richard M. McDermid $^{9}$, Raffaella Morganti$^{10,11}$,  Thorsten Naab$^{12}$, Tom Oosterloo$^{10,11}$, Marc Sarzi$^{13}$,Nicholas Scott$^5$, Paolo Serra$^{10}$, Anne-Marie Weijmans$^{14}$ and Lisa M. Young$^{15}$
\affil{$^1$Observatoire de Paris, France; $^2$ESO, Garching, Germany; $^3$CEA, Paris-Saclay, France; $^4$University of California, USA; $^5$University of Oxford, UK; $^6$Leiden University, the Netherlands; $^7$MPE, Garching, Germany; $^8$ESO, Garching, Germany; $^9$Gemini Observatory, Hilo, USA; $^{10}$ASTRON, Dwingeloo, The Netherlands; $^{11}$University of Groningen, The Netherlands; $^{12}$MPA, Garching, Germany; $^{13}$University of Hertfordshire, UK; $^{14}$University of Toronto, Canada; $^{15}$New Mexico Institute of Mining and Technology, Socorro, USA}
}

\begin{abstract}
We study the formation of early-type galaxies (ETGs) through mergers with a sample of 70 high-resolution numerical simulations of binary mergers of disc galaxies. These simulations encompass various mass ratios, initial conditions and orbital parameters. We find that binary mergers of disc galaxies with mass ratios of 3:1 and 6:1 are nearly always classified as Fast Rotators according to the ATLAS$^{\rm 3D}$ criterion: they preserve the structure of the input fast rotating spiral progenitors. Major disc mergers (mass ratios of 2:1 and 1:1) lead to both Fast and Slow Rotators. Most of the Slow Rotators hold a stellar Kinematically Distinct Core (KDC) in their 1-3 central kilo-parsec: these KDCs are built from the stellar components of the progenitors. The mass ratio of the progenitors is a fundamental parameter for the formation of Slow Rotators in binary mergers, but it also requires a retrograde spin for the progenitor galaxies with respect to the orbital angular momentum. The importance of the initial spin of the progenitors is also investigated in the library of galaxy mergers of the GalMer project.
\end{abstract}

\section{Introduction}
Numerical simulations have clearly shown that the global characteristics of the remnants of binary mergers between two equal-mass spiral galaxies, called major mergers, resemble those of Early-Type Galaxies (\textit{i.e} ellipticals \& lenticulars), hereafter ETGs \citep{1992ApJ...393..484B}. Modern work tends to quantify in detail the properties of major and minor merger remnants, together with thorough comparisons with observed properties of ETGs \citep{2005A&A...437...69B}.

In this paper, we further examine the role of the initial conditions (mass ratios, impact parameters, relative velocities, inclinations and spins of the progenitors) on the morphology and the kinematics of the remnants of binary galaxy mergers. We build two-dimensional momentum (intensity, velocity and velocity dispersion) maps of the merger remnants and analyse their apparent properties, directly linked with their orbital structures \citep{2005MNRAS.360.1185J}. Using two-dimensional maps enables us to compare our merger remnants directly with the complete, volume-limited sample of 260 local ETGs from the ATLAS$^{\rm 3D}$ survey \citep{2011MNRAS.413..813C}.

\section{Sample of simulations}
\begin{figure}
  \includegraphics[width=\columnwidth]{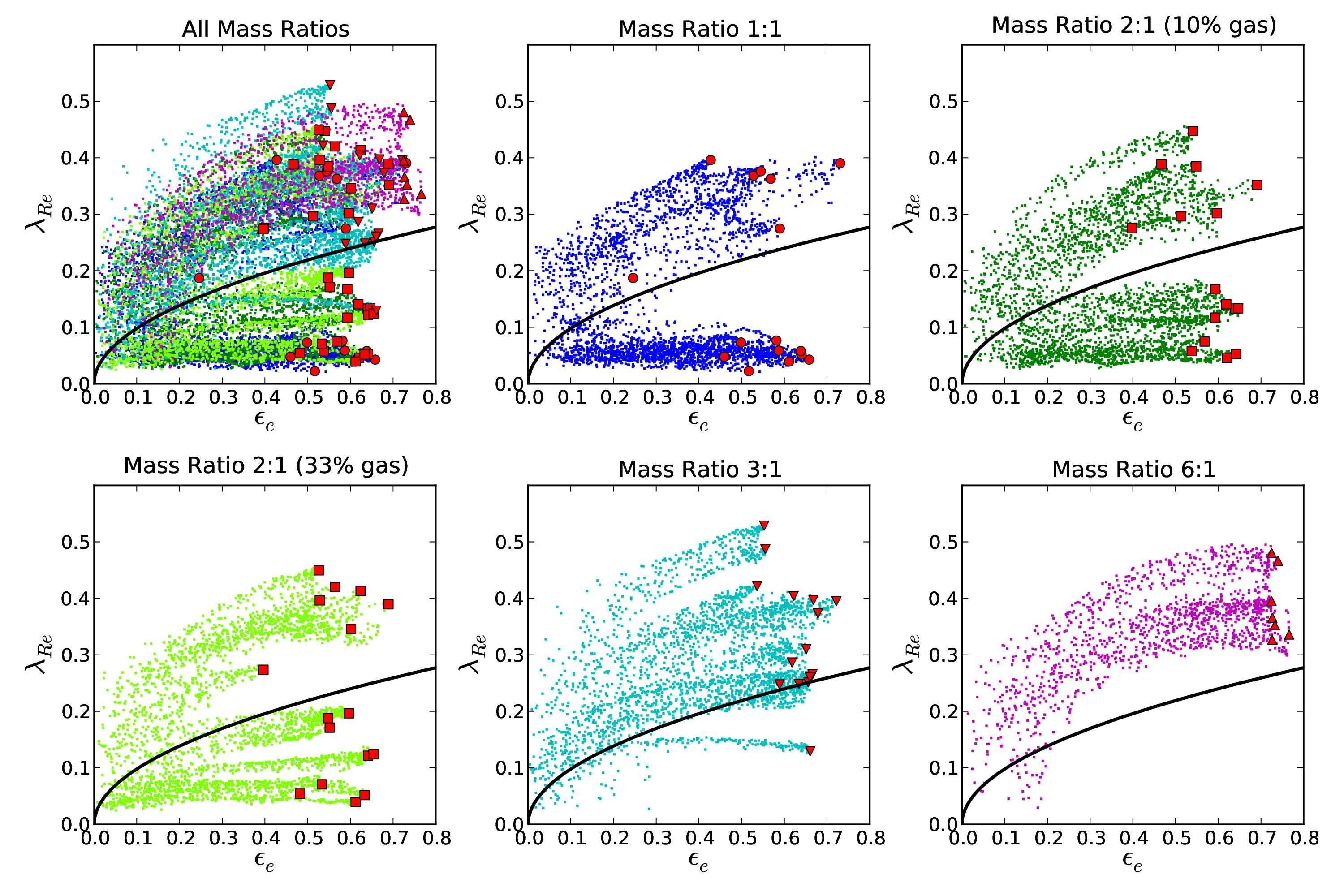}
  \caption{$\lambda_R - \epsilon$ diagram for all simulations of binary mergers of disc galaxies. The different mass ratios are labelled on top of each panels. The red symbols are for the projection which maximizes the ellipticity for a given remnant. The limit defining the slow/fast categories from ATLAS$^{\rm 3D}$ is plotted as the solid black line.}
  \label{fig1}
\end{figure}

We use the particle-mesh code described in \citet{2008MNRAS.389L...8B} and references therein. The softening lenght is 58~pc and the number of particles is 2~$\times$~10$^6$ for each components (gas, stars and dark matter) for each galaxies. This resolution is required to resolve properly the fluctuations of the gravitational potential during the merger, \textit{i.e} violent relaxation, which can significantly impact the morphology and kinematics of merger remnants \citep{2010MNRAS.406.2405B}.

We have simulated binary mergers of spiral galaxies with mass ratios of 1:1, 2:1, 3:1 and 6:1. The first progenitor, which is defined as the most massive for unequal-mass mergers, has a baryonic mass of 1.3~$\times$~10$^{11}$~M$_{\odot}$. The bulge fraction is $B/T = 0.20$ and the gas fraction in the disc is usually 10 per cent (33 per cent in some 2:1 mergers). This initial galaxy is representative for a \textit{Sb} spiral galaxy. The other progenitor has its total mass determined by the mass ratio and a gas fraction of 10 per cent. The main difference with the first progenitor is the lower bulge fraction with $B/T = 0.12$. This second progenitor galaxy is denoted as the \textit{Sc} spiral progenitor. 

We simulate 70 galaxy mergers with different initial conditions. We vary five parameters to study their importance on the properties of the final remnant: the mass ratio, the inclination of the spiral \textit{wrt} the orbital plane, the impact parameter, the relative velocity of the spirals and the orientation of the spin of the spirals. The spin of the spirals can be then either direct (parallel) or retrograde (anti-parallel) \textit{wrt} the orbital spin.

\section{Populations of fast and slow rotators}
\begin{figure}
  \includegraphics[width=0.5\columnwidth]{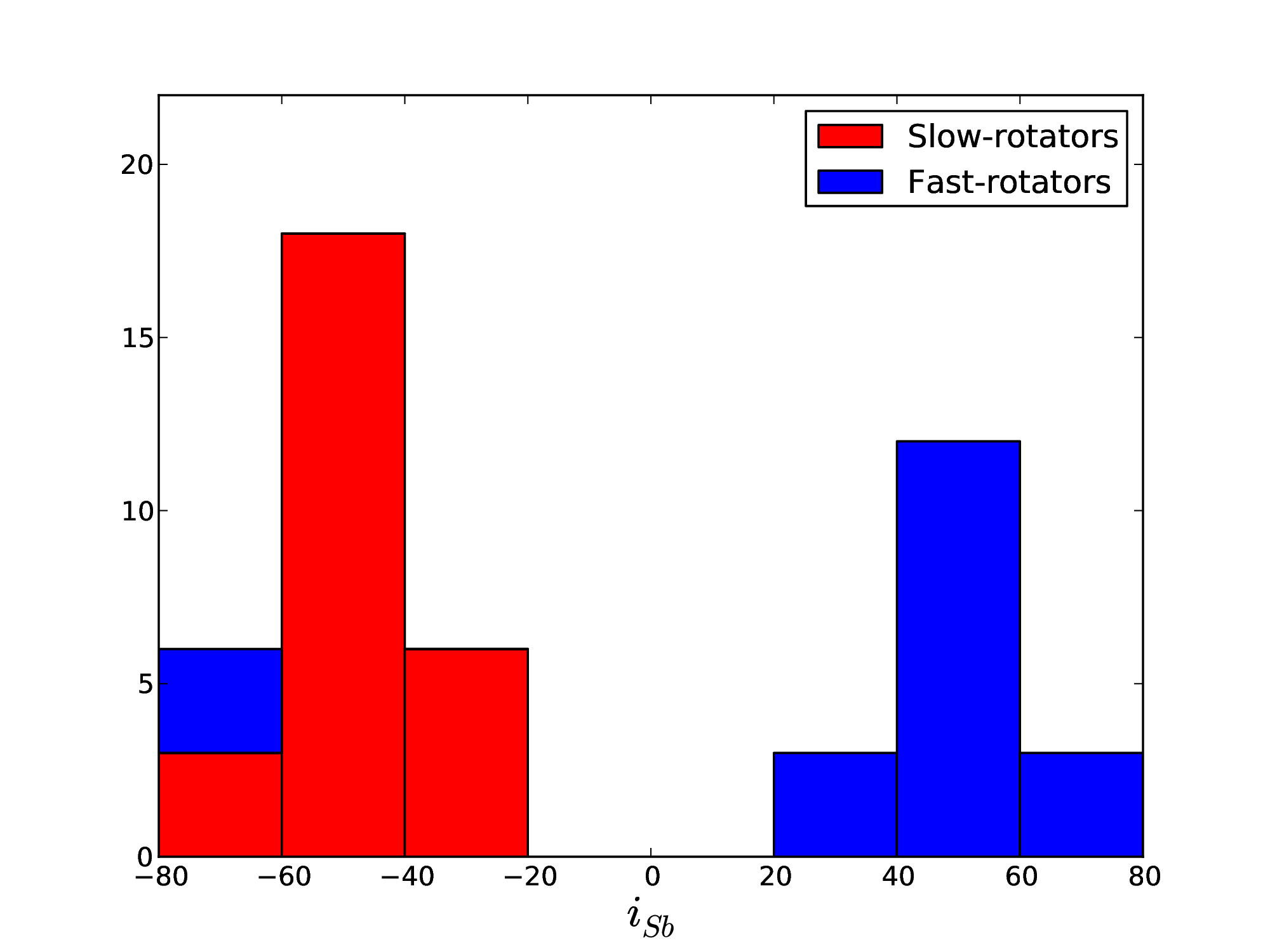}
  \includegraphics[width=0.5\columnwidth]{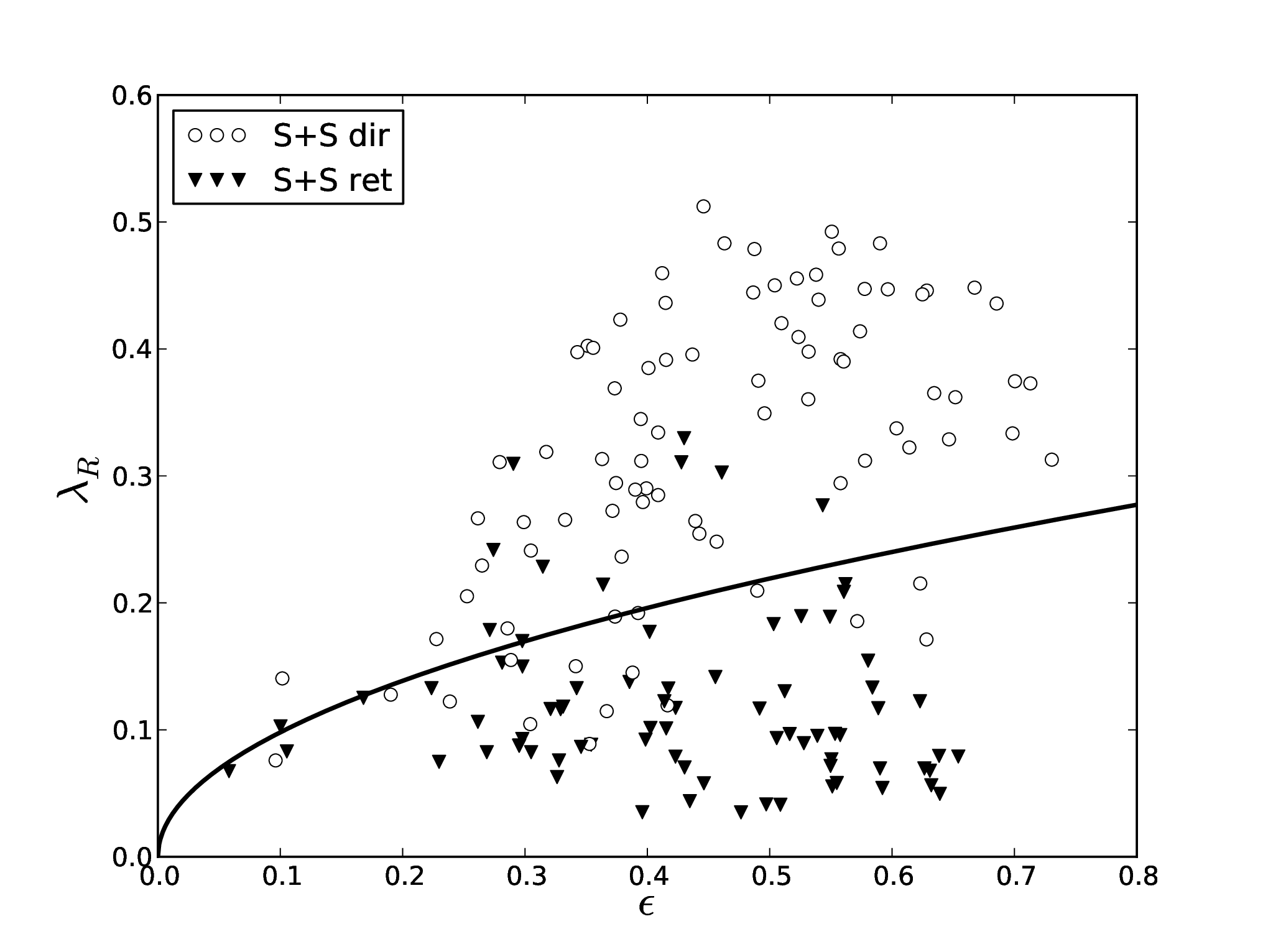}
  \caption{\textbf{Left:} Fraction of slow/fast rotators as a function of the inclination of the main progenitor with respect to the merger orbital plane for mass ratio 1:1, 2:1g10, 2:1g33. The negative inclination occurs when the spin of the Sb progenitor and the spin of the orbital angular momentum are anti-parallel. \textbf{Right:} $\lambda_R - \epsilon$ diagram for the GalMer merger remnants. In white symbols the spiral progenitors have a direct spin, in black symbols the spirals have a retrograde spin \textit{wrt} the orbital spin.}
  \label{fig2}
\end{figure}

To quantify the global kinematics of each system, using the velocity and velocity dispersion maps, we measure the $\lambda_R$ parameter, which is a robust proxy for the stellar projected angular momentum defined in \citet{2007MNRAS.379..401E}. This parameter is used to separate ETGs in two families: the fast rotators and the slow rotators \citep[see][for detailed properties on these families]{2011MNRAS.414.2923K,2011MNRAS.414..888E}. To quantify the morphology, we measure the ellipticity $\epsilon = 1 -b/a$ where $a$ and $b$ are the semi major- and minor-axes, respectively. To compute these two parameters, we project our simulated merger remnants over 200 differents viewing angles to obtain statistically representative results.

\subsection{High-resolution simulations}
Figure~\ref{fig1} shows the $\lambda_R - \epsilon$ diagram for the 70 high-resolution simulations of binary mergers for their 200 projections. All projected galaxies which are above (\textit{resp.} below) the black line are classified as fast rotators (\textit{resp.} slow rotators). We first note that almost all the merger remnants formed with a 3:1 or 6:1 are classified as fast rotators. They preserved the structure of the input fast rotating spiral progenitors. Major spiral mergers (with mass ratios of 1:1 and 2:1) lead to both fast and slow rotators. A look at their velocity maps reveals that most of the slow rotators hold a stellar Kinematically Distinct Core (KDC) in their 1-3 central kilo-parsec: these KDCs are built from the stellar components of the progenitors. The mass ratio is then a fundamental parameter for the formation of slow rotators in binary mergers.

The left panel of Figure~\ref{fig2} shows the number of fast and slow rotators for the 1:1 and 2:1 mergers as a function of the inclination of the \textit{Sb} spiral progenitor (a negative inclination occurs when the spin of the spiral is retrograde \textit{wrt} the orbital spin). The formation of the slow rotators requires a retrograde spin for the \textit{Sb} spiral with respect to the orbital angular momentum. Due to its massive bulge, the spin of the \textit{Sb} spiral is preserved during the merger while the \textit{Sc} spiral (with a lower bulge) always acquires the direct spin of the orbit. It thus forms a merger remnant with two contributions in counter-rotation and forms, in our simulations, an apparent KDC. These slow rotating merger remnants are then representative of the 2-$\sigma$ galaxies observed in the ATLAS$^{\rm 3D}$ survey \citep{2011MNRAS.414.2923K,2011MNRAS.414..888E}. The properties of these observed galaxies indicate that they are composed of two counter-rotating components and could have been formed via a single binary merger \citep[see also][]{crocker09}.

\subsection{The GalMer simulations}
The GalMer project consists of one of the largest publicly available sample of numerical simulations of interacting galaxies \citep{chili10}: with a high number of simulations but a relatively low resolution (a total number of particles of 1.2~$\times$~10$^5$ and a softening length of 280~pc).

To confirm the importance of the spins' orientation of the progenitors, we have selected in the GalMer database the simulations of 1:1 spiral~--~spiral mergers. The right panel of Figure~\ref{fig2} shows the $\lambda_R - \epsilon$ diagram for the edge-on view of the merger remnants. This plot confirms our previous findings: merging spirals with a retrograde orientation \textit{wrt} the orbital spin produces a remnant with a lower angular momentum content. However, there is a large scatter in the $\lambda_R$ values of these simulated galaxies. This is probably due to the lower resolution of the simulations: a look at the velocity maps of the merger remnants reveals a high level of numerical noise.


\begin{thebibliography}
\bibitem[\protect\citeauthoryear{Barnes}{1992}]{1992ApJ...393..484B} Barnes J.~E, 1992, ApJ, 393, 484  
\bibitem[\protect\citeauthoryear{Bois et al.}{2010}]{2010MNRAS.406.2405B} Bois M., et al., 2010, MNRAS, 406, 2405
\bibitem[\protect\citeauthoryear{Bournaud et al.}{2005}]{2005A&A...437...69B} Bournaud F., Jog C.~J., Combes F., 2005, A\&A, 437, 69
\bibitem[\protect\citeauthoryear{Bournaud et al.}{2008}]{2008MNRAS.389L...8B} Bournaud F., Duc P.-A., Emsellem E., 2008, MNRAS, 389, 8
\bibitem[\protect\citeauthoryear{Cappellari et al.}{2011}]{2011MNRAS.413..813C} Cappellari M., et al., 2011, MNRAS, 413, 813
\bibitem[\protect\citeauthoryear{Chilingarian et al.}{2010}]{chili10} Chilingarian I.~V., et al., 2010, A\&A, 518, 61 
\bibitem[\protect\citeauthoryear{Crocker et al.}{2009}]{crocker09} Crocker A.~F., et al., 2009, MNRAS, 393, 1255
\bibitem[\protect\citeauthoryear{Emsellem et al.}{2007}]{2007MNRAS.379..401E} Emsellem E., et al., 2007, MNRAS, 379, 401
\bibitem[\protect\citeauthoryear{Emsellem et al.}{2011}]{2011MNRAS.414..888E} Emsellem E., et al., 2011, MNRAS, 414, 888 
\bibitem[\protect\citeauthoryear{Jesseit et al.}{2005}]{2005MNRAS.360.1185J} Jesseit R., Naab T., Burkert A., 2005, MNRAS, 360, 1185
\bibitem[\protect\citeauthoryear{Krajnovi{\'c} et al.}{2011}]{2011MNRAS.414.2923K} Krajnovi{\'c} D., et al., 2011, MNRAS, 414, 2923 


\end{thebibliography}

\end{document}